\documentclass[aps,prl,showpacs,twocolumn,superscriptaddress]{revtex4-1}

\usepackage{graphicx} 
\usepackage{color}
\usepackage{array}
\usepackage{mathrsfs}
\usepackage{ulem}

\begin{document}

\title{Magnetic phase diagrams and magnetization plateaus of the spin-1/2 antiferromagnetic Heisenberg model on square-kagome lattice with three nonequivalent exchange interactions}

\author{Katsuhiro Morita}
\email[e-mail:]{katsuhiro.morita@rs.tus.ac.jp}
\affiliation{Department of Applied Physics, Tokyo University of Science, Tokyo 125-8585, Japan}

\author{Takami Tohyama}
\affiliation{Department of Applied Physics, Tokyo University of Science, Tokyo 125-8585, Japan}

\date{\today}

\begin{abstract}
Magnetization plateaus in quantum spin systems emerge in two-dimensional frustrated systems such as a kagome lattice. The spin-1/2 antiferromagnetic Heisenberg model on a square-kagome lattice is also appropriate for the study of the magnetization plateau. Motivated by recent experimental findings of such a square kagome lattice with three nonequivalent bonds, we investigate the phase diagrams and magnetization plateaus of the lattice using the exact diagonalization method. In addition to the previously reported 1/3 and 2/3 plateaus in the model with two equivalent bonds, we find a new 2/3 plateau whose magnetic structure is characterized by spontaneously broken four-fold rotational symmetry. The plateau appears only in the case of three nonequivalent bonds. We propose the possibility of finding plateaus including the new one.
\end{abstract}


\maketitle
\maketitle
Two-dimensional frustrated quantum spin systems give rise to novel quantum states such as quantum spin liquids and valence bond crystal (VBC) states owing to the influence of quantum fluctuation and frustration.
Magnetization plateaus provide a good playground for such novel quantum states.
In spin-1/2 two-dimensional frustrated Heisenberg models, magnetization plateaus in magnetic fields have theoretically been predicted in the kagome lattice (KL)~\cite{kagome1,kagome2,kagome3}, triangular lattice~\cite{tri1,tri-J1-J2}, $J_1$-$J_2$ square lattice~\cite{tri-J1-J2,J1-J21,J1-J22}, Shastry-Sutherland lattice~\cite{SS1}, checkerboard lattice~\cite{checker1,checker2}, and square-kagome lattice (SKL)~\cite{sqk1,sqk2,sqk3}. 
However, magnetization plateaus have been observed experimentally only for the triangular~\cite{triex1,triex2} and Shastry-Sutherland lattices~\cite{SSex1,SSex2}.
This is because of the difficulty in synthesizing compounds that fit to theoretical models.
In particular, many compounds exhibit lattice distortions that do not correspond to theoretical models~\cite{dskago1,dskago2,dskago3,dskago4,dskago5,dskago6,dskago7,dskago8,dskago9,dskago10,dskago11,dskago12,dssq1,dssq2,dssq3}. 
It is thus important to theoretically study lattices with distortions to make comparison easy with real compounds. 

\begin{figure}[tb]
  \begin{center}
\includegraphics[width=86mm]{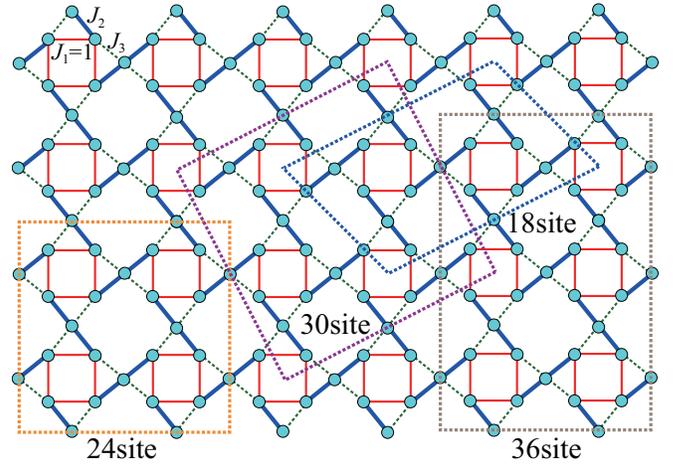}
\caption{(Color online) Lattice structure of SKL with three exchange interactions.  The red solid, blue thick, and green dashed lines denote the exchange interactions $J_1$, $J_2$, and $J_3$, respectively. We set $J_1=1$.
The shape of the system size $N$ with periodic boundary conditions is denoted by the blue dashed lines for $N=18$, the orange dashed lines for $N=24$, the purple dashed lines for $N=30$, and the gray dashed lines for $N=36$.
\label{site}}
  \end{center}
\end{figure}

Very recently, a compound with the spin-1/2 SKL with three nonequivalent exchange interactions, $J_1$, $J_2$, and $J_3$,  (see Fig.~\ref{site}) was synthesized~\cite{sqkcom}.
This compound will be a good candidate for exhibiting possible plateaus with the novel quantum state.
However, the SKL with the three nonequivalent exchange interactions has not been studied theoretically, and thus, their magnetic properties are unknown.
Only the SKL with $J_2=J_3$ (see Fig.~\ref{site}), i.e., the lattice with the two equivalent bonds, has been studied and two kinds of 1/3 plateaus with up-up-down (UUD) structure and VBC as well as a 2/3 plateau with VBC have been reported~\cite{sqk3}.
Therefore, it is necessary to investigate the ground state of the SKL with nonequivalent exchange interactions together with magnetization plateaus for the sake of forthcoming experiments.

In this letter, we study the ground states of the spin-1/2 Heisenberg model on the SKL with the three nonequivalent exchange interactions at zero temperature using the Lanczos-type exact diagonalization method. We obtain magnetic phase diagrams at zero magnetic field and finite magnetic fields inducing both 1/3 and 2/3 plateaus. 
We find a new 2/3 plateau with the magnetic structure that breaks four-fold rotational symmetry spontaneously. The origin of the plateau is attributed to the presence of the three nonequivalent exchange interactions.

\begin{figure*}[tb]
  \begin{center}
\includegraphics[width=176mm]{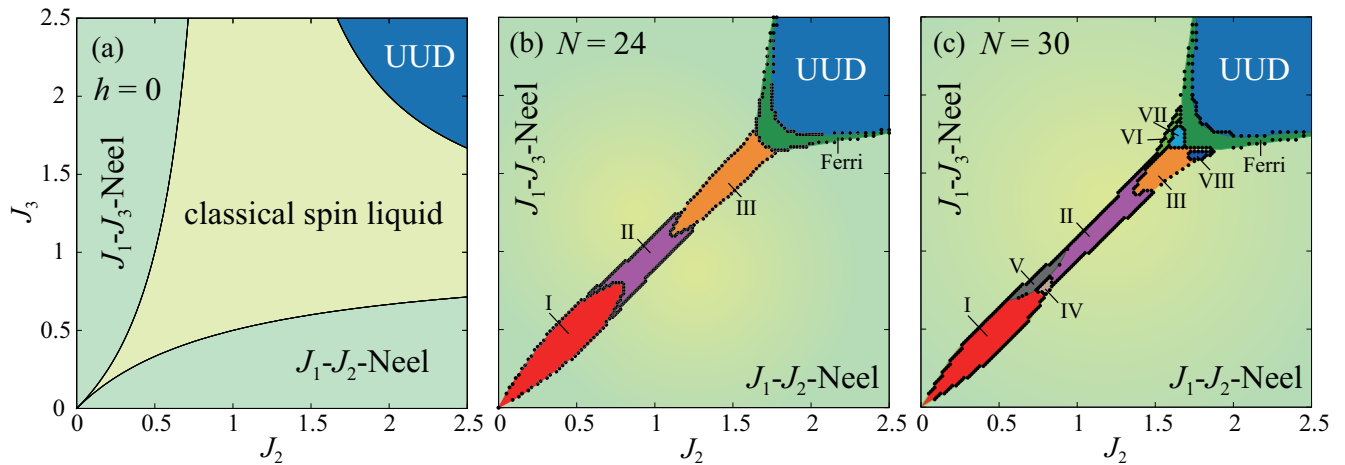}
\caption{(Color online) Phase diagram of the SKL Heisenberg model at $h=0$. (a) Classical phase diagram. Spin-1/2 quantum case for (b) $N=24$ and (c) $N=30$. The details of the phases are explained in the text.
\label{0field}}
  \end{center}
\end{figure*}

The Hamiltonian for the spin-1/2 SKL with the three exchange interactions in the magnetic field is defined as
\begin{eqnarray} 
H &=& \sum_{\langle i,j \rangle }J_{i,j} \mathbf{S}_i \cdot \mathbf{S}_j - h\sum_i S^{z}_i,
\end{eqnarray}
where $\mathbf{S}_i$ is the spin-$\frac{1}{2}$ operator at site $i$, $\langle i,j \rangle$ runs over the nearest-neighbor spin pairs, $J_{i,j}$ corresponds to one of $J_1$, $J_2$, and $J_3$ shown in Fig.~\ref{site}, and $h$ is the magnitude of the magnetic field in the z-direction.  In the following we set $J_{1}=1$ as the energy unit.
We perform the Lanczos-type exact diagonalization calculations at zero temperature for the SKL with the system size $N = 18, 24, 30$, and 36, as shown in Fig.~\ref{site}, under the periodic boundary conditions.

We first show the classical phase diagram of the ground states at $h=0$ in Fig.~\ref{0field}(a). 
The classical spin liquid in the phase diagram means that the ground state exhibits macroscopic degeneracy, degrees of which are more than those of the degrees of global spin rotation symmetry.
The $J_1$-$J_{2(3)}$-Neel phase has a magnetic structure in which the nearest-neighbor spins on plaquettes formed by the $J_1$ bonds align in an antiparallel manner, and spins connected to the plaquettes through the $J_{2(3)}$ bond are also antiparallel to the neighboring spins on the plaquettes.
On the other hand, in the UUD phase, all spins on the plaquettes are ferromagnetically arranged and the remaining spins are antiparallel to the plaquette spins.
We note that,  in contrast to the classical spin liquid phase, the degrees of degeneracy in the $J_1$-$J_{2(3)}$-Neel and UUD phases are determined by the global spin rotation symmetry.

Quantum effects change the phase diagram.
Figures~\ref{0field}(b) and \ref{0field}(c) show the phase diagram obtained by applying the Lanczos-type exact diagonalization for the $N=24$ and $N=30$ periodic lattices, respectively. The phase boundaries are determined by investigating the level crossing of the ground state and an excited state as well as the second derivative of the ground energy with respect to $J_2$ or $J_3$.
Along the $J_2=J_3$ line, we find five phases, I, II, III, Ferri, and UUD for $N=24$, and six phases, I, II, III, IV, Ferri, and UUD for $N=30$.
The detailed magnetic structures in the I, II, III, and IV phases have not been identified yet; however, the presence of  the phases are consistent with the previous report~\cite{sqk3}.
The UUD phase is the same as the classical one.
In the Ferri phase, the total spin $S$ of the ground states satisfies the condition $0 < S < \frac{1}{3}M_{\rm sat}$, where $M_{\rm sat}$ is the saturation magnetization.
We note that the UUD phase corresponds to the $M=1/3$ Ferri phase in the previous report~\cite{sqk3}.
Around the $J_2=J_3$ line for $N=30$, we find phases denoted by V, VI, VII, and VIII in Fig.~\ref{0field}(c). The phases are not well characterized because of the possibility of finite-size effects. In fact, the IV-VIII phases are not found for $N=24$. Moreover, in the thermodynamic limit, the phase diagram must be symmetric with respect to the $J_2=J_3$ line; however, it is not in Fig.~\ref{0field}(c). 
This is due to the geometry of the $N=30$ system, where 
there is no mirror symmetry between the $J_2$ and $J_3$ bonds (see Fig.~\ref{site}). 
On the other hand, the $N=18$, 24, and 36 systems have mirror symmetry, making them symmetric with respect to the $J_2=J_3$ line in the phase diagram, as shown in Fig.~\ref{0field}(b).

\begin{figure}[tb]
  \begin{center}
\includegraphics[width=86mm]{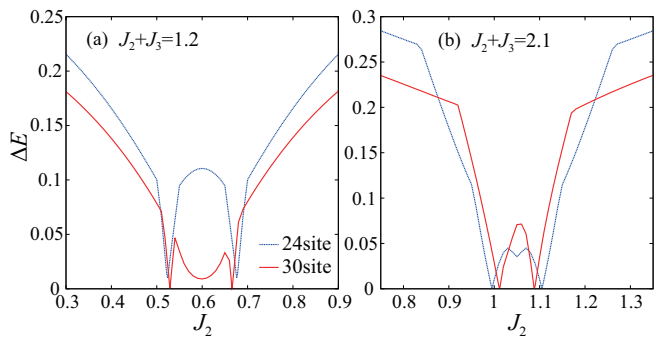}
\caption{(Color online) The energy gap $\Delta E$ between the first excited and ground states for $N=24$ and 30. (a) The $J_2+J_3=1.2$ line. (b) The $J_2+J_3=2.1$ line.
\label{de}}
  \end{center}
\end{figure} 

To confirm the presence of I and II phases in the thermodynamic limit,  we calculate the energy gaps, $\Delta E$, between the first excited and ground states for $N=24$ and  $N=30$.  Figures~\ref{de}(a) and \ref{de}(b) show $\Delta E$ on the $J_2+J_3=1.2$ line crossing the I phase and the $J_2+J_3=2.1$ line crossing the II phase, respectively.
The line segment on the $J_2$ axis joining the two $\Delta E=0$ or minimum points in Fig.~\ref{de}(a) [\ref{de}(b)] corresponds to the I [II] phase.
When $N$ increases from 24 to 30, the width of the I and II phases becomes narrow. Therefore, there is a possibility that these phases exist only on the $J_2=J_3$ line in the thermodynamic limit. In the vicinity of the II phase, $\Delta E$ become larger for $N=30$ than for $N=24$ [see Fig.\ref{de}(b)]. This suggests that the gap becomes finite in the thermodynamic limit. There is a possibility that quantum ground states such as quantum spin liquid or VBC states exist.
Further away from the $J_2=J_3$ line in Fig.~\ref{0field}(b) and \ref{0field}(c), we find the $J_1$-$J_{2(3)}$-Neel phase similar to the classical result; however, we cannot identify spin-liquid ground states for the present systems.  
In order to confirm the existence of quantum spin liquid and VBC phases at $h=0$, we need more detailed studies using larger size systems. This remains a future problem.

\begin{figure}[tb]
  \begin{center}
\includegraphics[width=86mm]{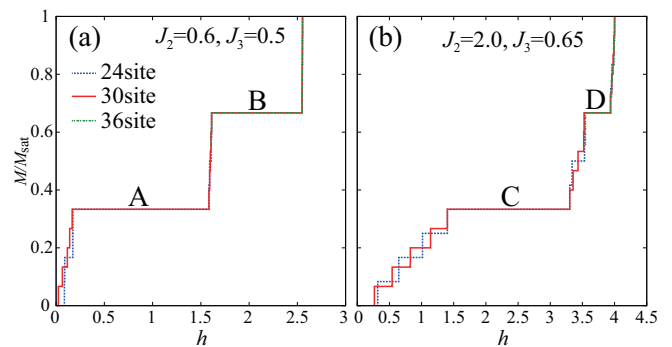}
\caption{(Color online) Magnetization curve of the SKL at zero temperature for $N=24$, 30, and 36. (a) $J_2=0.6$ and $J_3=0.5$, and (b) $J_2=2.0$ and $J_3=0.65$. There are the $1/3$ ($2/3$) plateaus in both (a) and (b), labeled by A and C (B and D), respectively. The D plateau is newly discovered in this work.
\label{m-h}}
  \end{center}
\end{figure}

Next, we investigate the ground states of the spin-1/2 SKL in the magnetic field.
Figures~\ref{m-h}(a) and \ref{m-h}(b) show the magnetization curve at zero temperature for $J_2=0.6$, $J_3=0.5$ and $J_2= 2.0$, $J_3=0.65$, respectively. 
The magnetization curves show the 1/3 and 2/3 plateaus denoted by A and C for $M/M_\mathrm{sat}=1/3$ and B and D for $M/M_\mathrm{sat}=2/3$. The width of these plateaus are almost size independent. Thus, these plateaus are expected to remain even in the thermodynamic limit.

\begin{figure}[!t]
  \begin{center}
\includegraphics[width=76mm]{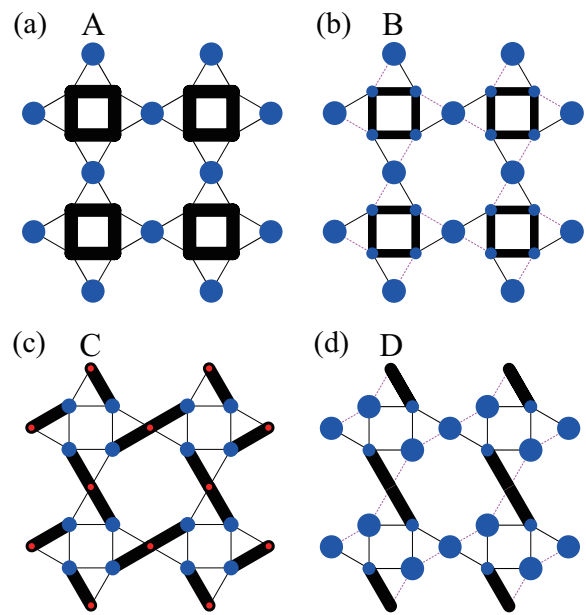}
\caption{(Color online) The nearest-neighbor spin correlation $\langle \mathbf{S}_i\cdot\mathbf{S}_j\rangle$ -$\langle S^z_i\rangle$$\langle S^z_j\rangle$ and the local magnetization $\langle S^z_i\rangle$. The plateaus, (a) A, (b) B, (c) C, and (d) D for $N=30$ shown in Fig.~\ref{m-h}. For D, a very small difference of $J_2$ between the two $J_2$ bonds is introduced (see text). Black solid (purple dashed) lines connecting two nearest-neighbor sites denote negative (positive) values of the spin correlation and their thicknesses represents the magnitudes of correlation. Blue (red) circles on each site denote positive (negative) value of $\langle S^z_i\rangle$ and their diameters represents its magnitude. We name A, B, C, and D the VBC-I, VBC-II, UUD, and AT phases, respectively.
\label{structure}}
  \end{center}
\end{figure}

The magnetic structures of the A, B, C, and D plateaus are shown in Figs~\ref{structure}(a), \ref{structure}(b), \ref{structure}(c), and \ref{structure}(d), respectively.
In A and B, the magnetic structure exhibits antiferromagnetically coupled four spins on the plaquette and almost fully polarized spins surrounding the plaquette.
Here, we name the phases of A and B, VBC-I and VBC-II, respectively.
The magnetic structure of C is the UUD structure, where the trimers whose central spin aligns downward and side-spins align upward are formed because of a large $J_2$, as shown in Fig~\ref{structure}(c).
The three magnetic structures, VBC-I, VBC-II, and UUD, have already been studied in the SKL with $J_2=J_3$~\cite{sqk3}. However, the D plateau has not been reported before, to the best of our knowledge. We will discuss the new plateau below.

In the D plateau for $N=24$ and $N=30$, there is no spontaneous symmetry breaking of the magnetic structure because of the finite-size effects. In order to observe the spontaneous symmetry breaking expected in the thermodynamic limit, we introduce a very small 0.01\% difference in $J_2$ between the two $J_2$ bonds, which corresponds to spatial anisotropy, and we calculate the magnetic structure of D for $N=30$, which is shown in Fig.~\ref{structure}(d). We note that the magnetic structure of D for the $N=36$ system is almost the same as Fig.~\ref{structure}(d), without the difference of $J_2$ because of the geometrical anisotropy of the system. (see Fig.~\ref{site}). We find that two kinds of trimers connected by the two $J_2$ bonds, i.e., a strongly connected antiferomagnetic trimer and a weakly connected ferromagnetic trimer, are ordered in D. This structure breaks the four-fold rotational symmetry spontaneously.
We name this phase the alternate trimerized (AT) phase.
The two kinds of trimers-ordered structures have been reported in a 2/3 plateau of the distorted diamond chain~\cite{DC}.
In the distorted diamond chain, translational symmetry is spontaneously broken and the period of the magnetic structure is doubled for the period of the lattice.
On the other hand, in the SKL, the translational symmetry does not break, but the four-fold symmetry spontaneously changes to two-fold symmetry.

\begin{figure}[tb]
  \begin{center}
\includegraphics[width=76mm]{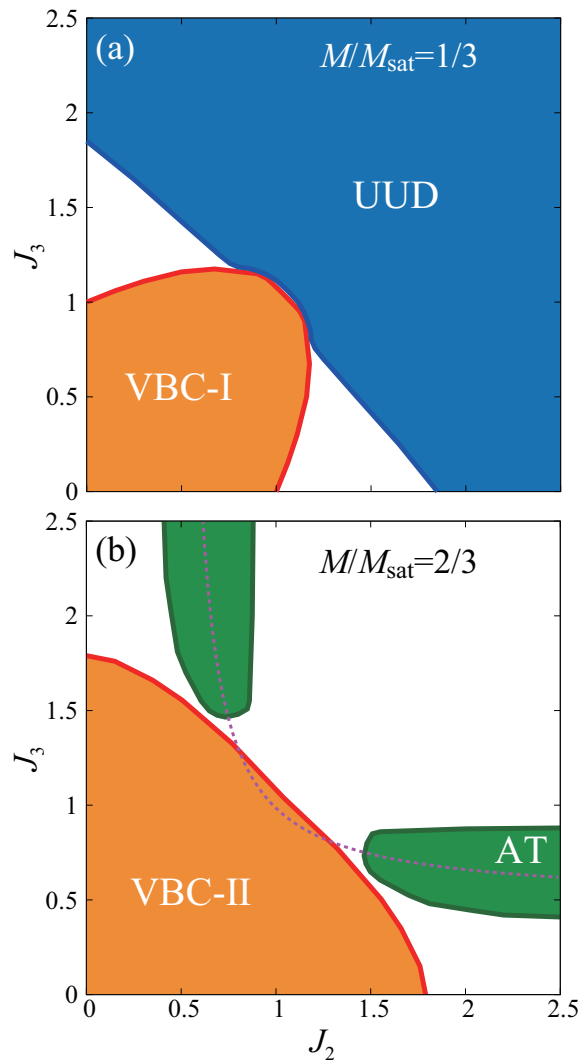}
\caption{(Color online) Phase diagrams of the SKL Heisenberg model at finite magnetization. (a) $M/M_\mathrm{sat}=1/3$ and (b) $M/M_\mathrm{sat}=2/3$. The regions where plateaus exist are colored by orange (VBC-I and VBC-II), blue (UUD), and green (AT). The purple dotted line in (b) represents the ground state phase boundary between a two-sublattice canted order and a four-sublattice canted order in the classical spin system.
\label{13-23}}
  \end{center}
\end{figure}

In order to clarify the robustness of the plateaus discussed above, we investigate the phase diagram for $M/M_\mathrm{sat}=1/3$ and 2/3. In order to obtain precise phase diagrams in the thermodynamic limit, we need to calculate the lower and upper magnetic fields, between which the $M/M_\mathrm{sat}=1/3$ or 2/3 state is the ground state for finite-size systems, and we have to perform a finite-size scaling of the fields. If the difference of the upper and lower field, $W$, is positive and finite in the thermodynamic limit, there is a 1/3 or 2/3 plateau at a given $J_2$ and $J_3$. To perform such a finite-size scaling, we use the exact diagonalization data for $N=18$, 24, and 30, where we choose the lattices that have a square shape, excluding  the less square-shaped $N=36$ lattice, and we perform a simple linear fitting of the data, assuming that the data are symmetric with respect to the exchange of $J_2$ and $J_3$. From the fitting and exploration to the thermodynamic limit, we judge a plateau to be present if $W>0$.

Figure~\ref{13-23}(a) shows the phase diagram for $M/M_\mathrm{sat}=1/3$.
The VBC-I phase corresponding to A in Fig.~\ref{m-h}(a) exists in the region of $J_2\lesssim J_1$ and $J_3\lesssim J_1$, since the VBC-I phase is characterized by strong antiferromagnetic correlations in the plaquettes controlled by $J_1$ as shown in Fig.~\ref{structure}(a). On the other hand, the UUD phase corresponding to C in Fig.~\ref{m-h}(b) covers the region of $J_3\gtrsim 2-J_2$, as expected from the magnetic structure of the UUD phase as shown in Fig.~\ref{structure}(c).
Figure~\ref{13-23}(b) shows the phase diagram for $M/M_\mathrm{sat}=2/3$. The VBC-II phase corresponding to B in Fig.~\ref{m-h}(a) exists in the region similar to the VBC-I phase for $M/M_\mathrm{sat}=1/3$. The AT phase appears in the regions of $0.5\lesssim J_2 (J_3) \lesssim1.0 $ and $1.5\lesssim J_3 (J_2)$. The presence of the AT phase at the regions is related to the phase boundary in the corresponding classical spin system. In fact, the AT phase is approximately located on the dotted line in Fig.~\ref{13-23}(b), given by the equation $J_3=J_2/(2J_2-1)$, which shows the ground state phase boundary between a two-sublattice canted order and a four-sublattice canted order in the classical system. The $z$ component of the spins in both orders does not break the four-fold rotational symmetry, which is different from the structure shown in the Fig.~\ref{structure}(d). However, since the ground states degenerate at the phase boundary, a state with the four-fold symmetry breaking can be a ground state just at the boundary. 
The quantum fluctuation on top of the degenerated classical ground state can lift the degeneracy due to the so-called order-by-disorder mechanism, resulting in the AT phase in these regions. Moreover, we confirm the existence of the AT phase even at $J_2=10.0$, $J_3=0.6$, out of the plotted region in Fig~\ref{13-23}(b). Therefore, the AT phase exists in wide regions.

In summary, inspired by the recent discovery of  a new spin-1/2 SKL compound with three nonequivalent exchange interactions~\cite{sqkcom}, we investigated the ground state of a spin-1/2 Heisenberg model on SKL with three nonequivalent exchange interactions in a magnetic field using the Lanczos-type exact diagonalization method.
We obtained phase diagrams at zero magnetic field and finite magnetic field inducing both 1/3 and 2/3 plateaus; we found four plateau phases. One of them is a new 2/3 plateau phase, which we name the AT phase, where the magnetic structure spontaneously breaks four-fold rotational symmetry. The origin of the plateau is attributed to the presence of three nonequivalent exchange interactions.
Since these four plateaus exist in wide regions for $J_2$ and $J_3$, the newly discovered compound is expected to exhibit the plateaus.
We hope that our study will motivate further experimental investigations on SKL compounds in the future.

\begin{acknowledgments}
This work was supported in part by MEXT as a social and scientific priority issue (Creation of new functional devices and high-performance materials to support next-generation industries) to be tackled by using post-K computer and by MEXT HPCI Strategic Programs for Innovative Research (SPIRE) (hp170274). The numerical calculation was partly carried out at the K Computer, Institute for Solid State Physics, The University of Tokyo, and the Information Technology Center, The University of Tokyo. 
\end{acknowledgments}

\end{document}